\documentclass[aps,prl,twocolumn,superscriptaddress,preprintnumbers]{revtex4-2}

\usepackage{graphicx}
\usepackage{dcolumn}
\usepackage{bm}
\usepackage{epstopdf}
\usepackage[caption=false]{subfig}

\usepackage{amsmath} 

\newcommand{\iso}[2]{\protect{\ensuremath{{}^{#1}\textrm{#2}}}\,}

\usepackage{verbatim}
\usepackage{xcolor}
\begin{document}
\preprint{LA-UR-21-28552}

\title{First Leptophobic Dark Matter Search From Coherent CAPTAIN-Mills} 

\affiliation{Bartoszek~Engineering,~Aurora,~IL~60506,~USA}
\affiliation{Columbia~University,~New~York,~NY~10027,~USA}
\affiliation{University~of~Edinburgh,~Edinburgh,~United~Kingdom}
\affiliation{Embry$-$Riddle~Aeronautical~University,~Prescott,~AZ~86301,~USA }
\affiliation{University~of~Florida,~Gainesville,~FL~32611,~USA}
\affiliation{Los~Alamos~National~Laboratory,~Los~Alamos,~NM~87545,~USA}
\affiliation{Massachusetts~Institute~of~Technology,~Cambridge,~MA~02139,~USA}
\affiliation{Universidad~Nacional~Aut\'{o}noma~de~M\'{e}xico,~CDMX~04510,~M\'{e}xico}
\affiliation{University~of~New~Mexico,~Albuquerque,~NM~87131,~USA}
\affiliation{New~Mexico~State~University,~Las~Cruces,~NM~88003,~USA}
\affiliation{Texas~A$\&$M~University,~College~Station,~TX~77843,~USA}

\author{A.A.~Aguilar-Arevalo}
\affiliation{Universidad~Nacional~Aut\'{o}noma~de~M\'{e}xico,~CDMX~04510,~M\'{e}xico}
\author{D.~S.\,M.~Alves}
\affiliation{Los~Alamos~National~Laboratory,~Los~Alamos,~NM~87545,~USA}
\author{S.~Biedron}
\affiliation{University~of~New~Mexico,~Albuquerque,~NM~87131,~USA}
\author{J.~Boissevain}
\affiliation{Bartoszek~Engineering,~Aurora,~IL~60506,~USA}
\author{M.~Borrego}
\affiliation{Los~Alamos~National~Laboratory,~Los~Alamos,~NM~87545,~USA}
\author{M.~Chavez$-$Estrada}
\affiliation{Universidad~Nacional~Aut\'{o}noma~de~M\'{e}xico,~CDMX~04510,~M\'{e}xico}
\author{A.~Chavez}
\affiliation{Los~Alamos~National~Laboratory,~Los~Alamos,~NM~87545,~USA}
\author{J.M.~Conrad}
\affiliation{Massachusetts~Institute~of~Technology,~Cambridge,~MA~02139,~USA}
\author{R.L.~Cooper}
\affiliation{Los~Alamos~National~Laboratory,~Los~Alamos,~NM~87545,~USA}
\affiliation{New~Mexico~State~University,~Las~Cruces,~NM~88003,~USA}
\author{A.~Diaz}
\affiliation{Massachusetts~Institute~of~Technology,~Cambridge,~MA~02139,~USA}
\author{J.R.~Distel}
\affiliation{Los~Alamos~National~Laboratory,~Los~Alamos,~NM~87545,~USA}
\author{J.C.~D'Olivo}
\affiliation{Universidad~Nacional~Aut\'{o}noma~de~M\'{e}xico,~CDMX~04510,~M\'{e}xico}
\author{E.~Dunton}
\affiliation{Columbia~University,~New~York,~NY~10027,~USA}
\author{B.~Dutta}
\affiliation{Texas~A$\&$M~University,~College~Station,~TX~77843,~USA}
\author{A.~Elliott}
\affiliation{Embry$-$Riddle~Aeronautical~University,~Prescott,~AZ~86301,~USA }
\author{D.~Evans}
\affiliation{Los~Alamos~National~Laboratory,~Los~Alamos,~NM~87545,~USA}
\author{D.~Fields}
\affiliation{University~of~New~Mexico,~Albuquerque,~NM~87131,~USA}
\author{J.~Greenwood}
\affiliation{Embry$-$Riddle~Aeronautical~University,~Prescott,~AZ~86301,~USA }
\author{M.~Gold}
\affiliation{University~of~New~Mexico,~Albuquerque,~NM~87131,~USA}
\author{J.~Gordon}
\affiliation{Embry$-$Riddle~Aeronautical~University,~Prescott,~AZ~86301,~USA }
\author{E.~Guarincerri}
\affiliation{Los~Alamos~National~Laboratory,~Los~Alamos,~NM~87545,~USA}
\author{E.C.~Huang}
\affiliation{Los~Alamos~National~Laboratory,~Los~Alamos,~NM~87545,~USA}
\author{N.~Kamp}
\affiliation{Massachusetts~Institute~of~Technology,~Cambridge,~MA~02139,~USA}
\author{C.~Kelsey}
\affiliation{Los~Alamos~National~Laboratory,~Los~Alamos,~NM~87545,~USA}
\author{K.~Knickerbocker}
\affiliation{Los~Alamos~National~Laboratory,~Los~Alamos,~NM~87545,~USA}
\author{R.~Lake}
\affiliation{Embry$-$Riddle~Aeronautical~University,~Prescott,~AZ~86301,~USA }
\author{W.C.~Louis}
\affiliation{Los~Alamos~National~Laboratory,~Los~Alamos,~NM~87545,~USA}
\author{R.~Mahapatra}
\affiliation{Texas~A$\&$M~University,~College~Station,~TX~77843,~USA}
\author{S.~Maludze}
\affiliation{Texas~A$\&$M~University,~College~Station,~TX~77843,~USA}
\author{J.~Mirabal}
\affiliation{Los~Alamos~National~Laboratory,~Los~Alamos,~NM~87545,~USA}

\author{R.~Moreno}
\affiliation{Embry$-$Riddle~Aeronautical~University,~Prescott,~AZ~86301,~USA }
\author{H.~Neog}
\affiliation{Texas~A$\&$M~University,~College~Station,~TX~77843,~USA}
\author{P.~deNiverville}
\affiliation{Los~Alamos~National~Laboratory,~Los~Alamos,~NM~87545,~USA}
\author{V.~Pandey}
\affiliation{University~of~Florida,~Gainesville,~FL~32611,~USA}
\author{J.~Plata$-$Salas}
\affiliation{Universidad~Nacional~Aut\'{o}noma~de~M\'{e}xico,~CDMX~04510,~M\'{e}xico}
\author{D.~Poulson}
\affiliation{Los~Alamos~National~Laboratory,~Los~Alamos,~NM~87545,~USA}
\author{H.~Ray}
\affiliation{University~of~Florida,~Gainesville,~FL~32611,~USA}
\author{E.~Renner}
\affiliation{Los~Alamos~National~Laboratory,~Los~Alamos,~NM~87545,~USA}
\author{T.J.~Schaub}
\affiliation{University~of~New~Mexico,~Albuquerque,~NM~87131,~USA}
\author{M.H.~Shaevitz}
\affiliation{Columbia~University,~New~York,~NY~10027,~USA}
\author{D.~Smith}
\affiliation{Embry$-$Riddle~Aeronautical~University,~Prescott,~AZ~86301,~USA }
\author{W.~Sondheim}
\affiliation{Los~Alamos~National~Laboratory,~Los~Alamos,~NM~87545,~USA}
\author{A.M.~Szelc}
\affiliation{University~of~Edinburgh,~Edinburgh,~United~Kingdom}
\author{C.~Taylor}
\affiliation{Los~Alamos~National~Laboratory,~Los~Alamos,~NM~87545,~USA}
\author{W.H.~Thompson}
\affiliation{Los~Alamos~National~Laboratory,~Los~Alamos,~NM~87545,~USA}
\author{M.~Tripathi}
\affiliation{University~of~Florida,~Gainesville,~FL~32611,~USA}
\author{R.T.~Thornton}
\affiliation{Los~Alamos~National~Laboratory,~Los~Alamos,~NM~87545,~USA}
\author{R.~Van~Berg}
\affiliation{Bartoszek~Engineering,~Aurora,~IL~60506,~USA}
\author{R.G.~Van~de~Water}
\affiliation{Los~Alamos~National~Laboratory,~Los~Alamos,~NM~87545,~USA}
\author{S.~Verma}
\affiliation{Texas~A$\&$M~University,~College~Station,~TX~77843,~USA}
\author{K.~Walker}
\affiliation{Embry$-$Riddle~Aeronautical~University,~Prescott,~AZ~86301,~USA }

\collaboration{The CCM Collaboration}

\begin{abstract}
We report the first results of a search for leptophobic dark matter (DM) from the Coherent CAPTAIN-Mills (CCM) liquid argon (LAr) detector.  An engineering run with 120 photomultiplier tubes (PMTs) and $17.9 \times 10^{20}$ protons-on-target (POT) was performed in Fall 2019 to study the characteristics of the CCM detector.   The operation of this 10-ton detector was strictly light-based with a threshold of 50 keV and used coherent elastic scattering off argon nuclei to detect DM.  
%
Despite only 1.5 months of accumulated luminosity, contaminated LAr, and non-optimized shielding, CCM's first engineering run has already achieved sensitivity to previously unexplored parameter space of light dark matter (LDM) models with a baryonic vector portal. With an expected background of 115,005 events, we observe 115,005+16.5 events which is compatible with background expectations.  For a benchmark mediator-to-DM mass ratio of $m_{_{V_B}}/m_{\chi}=2.1$, DM masses within the range $9\,\text{MeV} \lesssim m_\chi \lesssim 50\,\text{MeV}$ are excluded at 90\% C.\,L. in the leptophobic model after applying the Feldman-Cousins test statistic.  CCM's upgraded run with 200 PMTs, filtered LAr, improved shielding, and ten times more POT will be able to exclude the remaining thermal relic density parameter space of this model, as well as probe new parameter space of other leptophobic DM models.
\end{abstract}

\maketitle





\paragraph{\textbf{Introduction}}
CCM is a proton fixed target experiment capable of detecting signals of new physics via coherent elastic nuclear scattering. As such, it can explore a variety of sub-GeV DM models interacting via light mediators. Since the dominant production and detection modes for DM occur via hadronic processes (such as pion decays and nuclear scattering), CCM has a unique sensitivity to leptophobic DM models \cite{Batell:2014yra,Coloma:2015pih,Dror:2017ehi,Berlin:2018bsc,Boyarsky:2021moj} that are difficult to probe in electron- or photon-beam experiments. LDM models that do not couple to electrons and photons tend to be unviable due to a lack of annihilation channels that could yield the correct thermal relic abundance. However, this issue is easily circumvented since loop processes will invariably generate a small mixing of the hadronic mediator with the Standard Model (SM) photon, opening viable parameter space where DM can efficiently annihilate in the early Universe \cite{Batell:2014yra}.

\begin{figure}[h]
   \centering
   \includegraphics[width=0.48\textwidth]{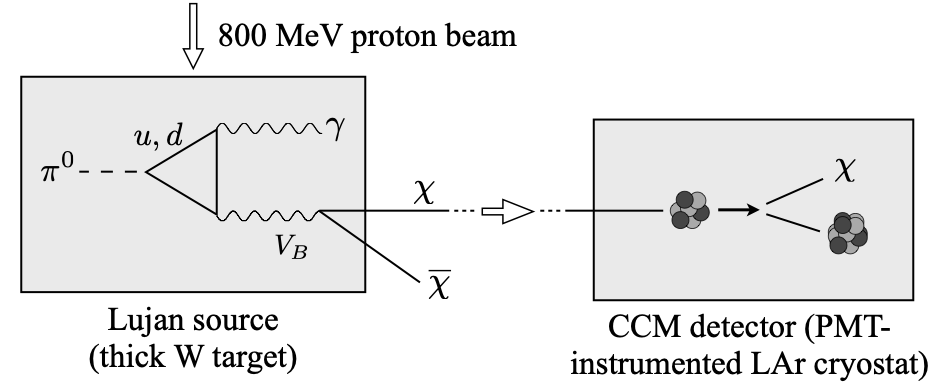}
   \caption{
   Leptophobic DM ($\chi$) is produced at the Los Alamos Neutron Science CEnter (LANSCE) through decays of light vector mediators ($V_{_B}$) coupled to baryon number, which in turn can be produced in rare pion decays. This generates a flux of DM particles that can coherently scatter off \iso{40}{Ar} nuclei in the CCM detector. The resulting nuclear recoil signals have a harder spectrum ($\gtrsim~50$~keV) than the Coherent Elastic Neutrino Nuclear Scatter (CE$\nu$NS) background from $\pi^+\to\mu^+\nu$ decays at rest.}
   \label{fig:LDMneutrinoScatter}
\end{figure}

CCM operates at the LANSCE Lujan Center where 800\,MeV protons are delivered at a rate of 20\,Hz in a 280\,ns triangular pulse from the LANSCE beamline and interact in a thick tungsten target~\cite{LISOWSKI2006910,Bultman:1998,MOCKO201327}, copiously producing $\pi^0$'s. Many light dark sector models predict rare pion decays to vector mediators that couple to quarks, $\pi^0\to \gamma V_{_B}$, as illustrated in Fig.\,\ref{fig:LDMneutrinoScatter} \cite{Batell:2014yra,Dobrescu:2014ita}. If kinematically allowed, these mediators can promptly decay to DM particles ($V_{_B}\to\chi\bar\chi$), where some of them travel to the CCM detector and scatter off nuclei through the same $V_{_B}$ mediator. CCM is designed to measure coherent scattering off \iso{40}{Ar} with an energy threshold of approximately 10 keV. 
The $A^2$-enhanced cross-section due to coherent scattering off nuclei, in this case Ar \,\cite{deNiverville:2015mwa,CCM:2021leg}, where $A$ is the number of nucleons in the nucleus, further improves CCM's sensitivity to DM masses below $m_{\pi^0}$.  As a result, CCM is able to extend the physics reach beyond the MiniBooNE experiment~\cite{Aguilar-Arevalo:2018wea,Aguilar-Arevalo:2017mqx} which explored similar models.
This letter describes the results of a search for leptophobic DM using the CCM120 coherent nucleon scattering data \cite{CCM:2021leg}.  Natural units ($\hbar=c=1$) are used throughout this letter.



 \paragraph{\textbf{Description of the CCM120 Detector}} CCM is an upright cylindrical stainless-steel detector with a diameter of 2.58\,m and a height of 2.26\,m.  It consists of a cryostat filled with 10\,tons of LAr, $\sim$3\,tons of active veto, and $\sim$5\,tons of fiducial volume. The 128\,nm scintillation light is produced from the decay of excited Ar dimer states, and observed by 120 \mbox{8-inch} R5912-mod2 Hamamatsu PMTs. Ninety-six PMTs are coated with Tetraphenyl Butadiene (TPB) to shift the scintillation light to visible light. Mylar foils painted with TPB cover the inner walls of the cryostat in between the PMTs to increase the light collection efficiency. The uncoated PMTs are used to disentangle the properties of the PMTs and LAr with that of the TPB. Both the coated and uncoated PMTs are used in the physics analysis.  Their performance characteristics are determined from the Optical Model\,\cite{CCM:2021leg}, a monte-carlo program using GEANT4, to determine the optical response of the CCM detector for the physics analysis.  The veto region is instrumented with twenty-three \mbox{1-inch} and five 8-inch PMTs that reject interactions occurring outside the detector.
The signals from the PMTs are read out by CAEN\,VX1730 500\,MHz digitizer boards in a 16\,$\mu$s window for each beam trigger,  random trigger, and LED trigger (calibration). Ten of the 16\,$\mu$s are used to measure the beam-out-of-time background.  Further experimental details can be found in Ref.\,\cite{CCM:2021leg}.
The upgraded follow-on experiment, CCM200~\cite{CCM:2021leg}, will include many improvements learned from CCM120, and will run for three years to significantly improve dark sector searches.

\paragraph{\textbf{Leptophobic DM Signal}} The leptophobic dark sector model considered in this analysis consists of a scalar DM candidate, $\chi$, and a vector portal communicating with the SM quarks via gauged baryon number \cite{Batell:2014yra,Coloma:2015pih,Dror:2017ehi,Berlin:2018bsc,Boyarsky:2021moj}. The interactions of the vector mediator, $V_{_B}$, are given by:
\begin{equation}
\mathcal{L}_B \supset -V_{_B}^\mu\,\left(
g_{_B} J_\mu^B\,+\,g_\chi J_\mu^\chi\,+\, \epsilon_{_B} e J_\mu^\text{EM}\right)\,
\end{equation}
where $J_\mu^B$ is the SM baryonic current, $J_\mu^B \equiv \frac{1}{3}\sum_i \bar q_i \gamma_\mu q_i$, and $J_\mu^\chi$ is the scalar DM current, $J_\mu^\chi \equiv i(\chi^*\partial_\mu\chi-\chi\partial_\mu\chi^*)$. The small coupling of $V_{_B}$ to the SM electromagnetic current $J_\mu^\text{EM}$ is naturally induced by loop effects, which generate a kinetic mixing $\epsilon_{_B}$ between $V_{_B}$ and the SM photon with the typical size of $\epsilon_{_B} \sim e g_{_B}/(4 \pi)^2$. This suppressed coupling has a negligible effect on DM production and detection at CCM; however, its presence is crucial to allow DM to annihilate in the early Universe efficiently, and it controls the DM relic abundance.

The other relevant parameters of this model are the DM mass, $m_\chi$, and the vector mediator mass, $m_{_{V_B}}$. For concreteness, our data analysis will set limits on a slice of parameter space where we fix the following parameters: (i) the mediator-to-DM mass ratio $m_{_{V_B}}/m_\chi=2.1$, (ii) the DM coupling strength to the vector portal $\alpha_\chi \equiv g_\chi^2/(4\pi) = 0.5$, and (iii) the $V_{_B}-\gamma$ kinetic mixing $\epsilon_{_B} = e g_{_B}/(4 \pi)^2$ to define the thermal relic density target.

The benchmark ratio of $m_{_{V_B}}/m_\chi =2.1$, in particular, captures a region of parameter space of baryonic vector portal models that remains viable. The dark matter annihilation rate in the early universe that sets the relic abundance, $\chi\chi^*\to V_{_B}^* \to e^+e^-$, scales as $\alpha_{_B} \alpha_\chi/ \big(m_{_{V_B}}^2-(2\,m_\chi)^2\big)^2$, where $\alpha_{_B} \equiv g_{_B}^2/(4\pi)$. When $m_{_{V_B}}\gg 2\,m_\chi$, the product of couplings needs to be sufficiently large, $\alpha_{_B} \alpha_\chi \sim 1.5\times10^{-11}\, (m_{_{V_B}}/\,\text{MeV})^4$,
 in order to avoid a dark matter {\it overabundance}. Unfortunately, this range of couplings is excluded by a combination of beam dump and fixed target experiments, including MiniBooNE and NA62.  On the other hand, when $m_{_{V_B}}$ is close to $2\,m_\chi$, the annihilation process is resonantly enhanced, and the couplings $\alpha_{_B} \alpha_\chi$ need to be proportionally suppressed to avoid a dark matter {\it underabundance}. Indeed, for our benchmark choices of $m_{_{V_B}}/m_\chi=2.1$ and $\alpha_\chi=0.5$, the values of $\alpha_{_B}$ yielding the correct dark matter relic density are sufficiently small and remain compatible with present experimental constraints in the range $2\, m_e \lesssim m_\chi \lesssim 13$\;MeV. 

The dominant process for DM production at the LANSCE Lujan source is given by $\pi^0$ production from the 800 MeV proton beam impinging on the W target, followed by rare decays of $\pi^0 \to \gamma\, (V_{_B}\to \chi\chi^*)$. Assuming $m_{_{V_B}} < m_{\pi^0}$, the branching ratio for this rare decay is given by \cite{Batell:2014yra}:
\begin{eqnarray}
    \frac{\mathrm{Br}(\pi^0 \to \gamma V_{_B})}{\mathrm{Br}(\pi^0\to\gamma\gamma)} &~=~& 2\left(\frac{g_{_B}}{e}-\epsilon_{_B}\right)^2\left(1-\frac{m_{_{V_B}}^2}{m_{\pi^0}^2}\right)^3\nonumber\\
    &~\simeq~&2\,\frac{\alpha_{_B}}{\alpha}\left(1-\frac{m_{_{V_B}}^2}{m_{\pi^0}^2}\right)^3\,,
\end{eqnarray}
where $\alpha\equiv e^2/(4\pi)$ is the fine-structure constant.

The DM signal, including production through $\pi^0 \to \gamma\, (V_{_B}\to \chi\chi^*)$, propagation to the CCM detector, and $V_{_B}$-mediated coherent nuclear scattering off of \iso{40}{Ar}, is simulated using the BdNMC event generator\,\cite{deNiverville:2016rqh}. 

\paragraph{\textbf{Data Analysis and Fits}} The details of the data analysis are given in Ref.\,\cite{CCM:2021leg}. It will suffice to say that individual pulses are found for each PMT waveform, and these pulses are combined to generate an accumulated waveform from which events are extracted. 
The length of the event is measured according to the activity within the detector and acts as a proxy for particle identification. A length cut of 48\,ns keeps 80\% of the simulated nuclear recoil events while rejecting 60\% of \iso{39}{Ar} events, a naturally occurring contaminant to \iso{40}{Ar} with a decay rate of $\sim$1 decay/sec/kg. The integrated charge for each event is measured over the length of the event.
Measurements of the proton on target $\gamma$-ray flash using an EJ-301 detector in a neighboring flight path provides a measurement for beam $T_0$, the time when the first speed-of-light particle (neutrino or assumed LDM) are expected to arrive at CCM. The time difference between $T_0$ and when CCM starts seeing the slower beam-related neutron interactions is 210\,ns, of  which the first 190\,ns is used as the signal Region of Interest (ROI) in the analysis to measure the signal rate.
The energy of the event is calibrated using a \iso{22}{Na} (2.2\,MeV) gamma-ray source located in the center of the detector. The detector response is $15.1\pm4.0$\,PEs/MeVee (electromagntic equivalent), and assumed to be linear to $>$200~keV and is consistent with simulations using the Optical Model.

The calibration data from the sources and laser runs are used to quantify the detector response model~\cite{CCM:2021leg}. Ten variables are identified and tuned in the process.  Using all the data, the absorption length for scintillation light in contaminated LAr is $\sim$50\;cm compared to 180\,cm in pure LAr \cite{ProtoDUNE:2021cuesta}. This drastically reduces the light output and is due to O$_2$ and H$_2$O contamination. 
The planned upgraded CCM200 detector will have a recirculation/filtration system for the LAr, which will reduce these sources of contamination~\cite{CCM:2021leg}.
Nuclear recoil events are simulated from 10\,keV to 5\,MeV to map out the detector's response. Using the shorter absorption length, the detection efficiency for a 100\,keV (1\,MeV) nuclear recoil event is 2\% (20\%).  The nuclear recoil efficiencies have not been validated with data; however, we plan to deploy a neutron source to better constrain the nuclear recoil efficiencies in the CCM200 detector simulation.

\begin{figure}[htp]
    \centering
    \includegraphics[width=0.48\textwidth]{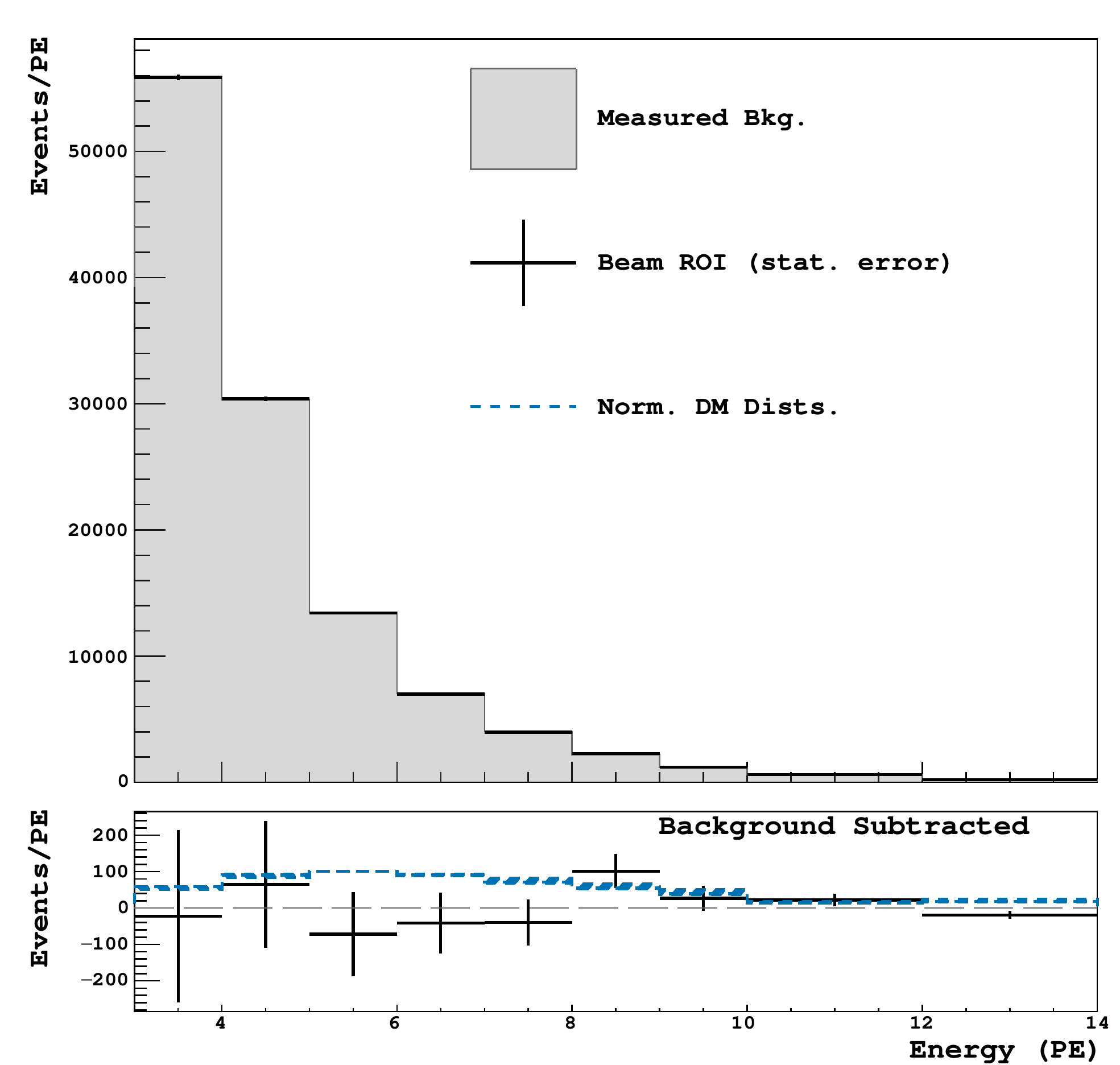}
    \caption{The top frame shows the reconstructed energy distribution after all cuts are applied. The background prediction based on the beam-out-of-time window is the shaded region, and measured data in the beam signal region of interest are the solid lines. The bottom frame shows the background subtracted distribution along with a blue line that is arbitrarily normalized to show the shape of the expected DM distribution.  The thickness of the blue line shows the variation due to 383 different $m_{_V}$, $m_{\chi}$ mass combinations.}
    \label{fig:dataDist}
\end{figure}

The CCM120 2019 engineering run accumulated $17.9\times10^{20}$\, POT, or  56,860,679 triggers, in a 1.5~month period. The signal event distribution and predicted background are shown in Fig.\,\ref{fig:dataDist}. The background is measured using the 4.180\,$\mu$sec beam-out-of-time window, which is 22 times bigger than the 190\,ns beam ROI. The beam-out-of-time window is split into 190\,ns time bins and no variations above statistical fluctuations are observed.
Since the background is measured directly from the data, the only systematic errors considered are on the prediction of the DM signal. A total fractional systematic error of 22.6\% is determined for reconstructed DM events, this includes the uncertainty in the amount of POT which is measured to an accuracy of 0.7\%, the uncertainty coming from the quenching factor of nuclear recoil events in LAr, and the uncertainty due to propagating the covariance matrix from the Optical Model parameters~\cite{CCM:2021leg}.
The background prediction is consistent with what is observed in the beam ROI.  The observed data includes 115,005 events, and after background subtraction there remains $16.5\pm338.4$ events, which is consistent with no observed excess. 

The $m_{_{V_B}}$-- $m_{\chi}$ parameter space is scanned within the intervals $m_{_{V_B}} = [0.3- 134]$ MeV and $m_\chi = [0.1-67]$ MeV to generate 90\% C.\,L. exclusion limits on $\alpha_{_B}$.
\begin{figure}[t]
    \centering
    \includegraphics[width=0.48\textwidth]{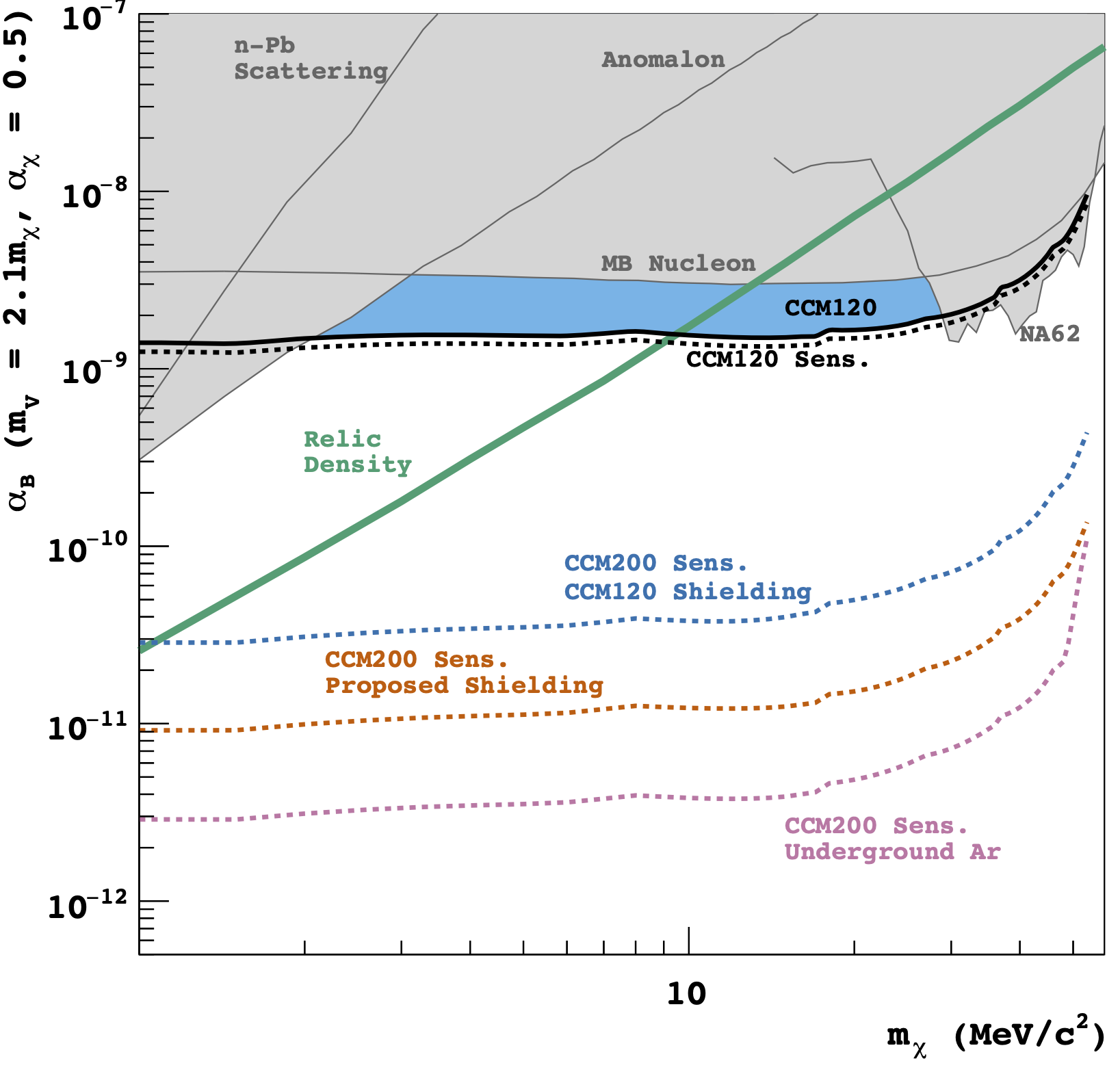}
    \caption{
    Median sensitivity (dotted black curve) and observed 90\% C.\,L. exclusions (solid black curve) set by CCM120 on the baryonic coupling strength of the vector portal mediator, $\alpha_{_B}$, as a function of the DM mass, $m_\chi$. The green line denotes the $m_\chi-\alpha_{_B}$ relation that predicts the observed DM thermal relic abundance, under the assumption of a small effective $V_{_B}-\gamma$ kinetic mixing of $\epsilon_{_B}=e\sqrt{\alpha_{_B}}/(4\pi)^{3/2}$. The shaded gray regions have been constrained by previous experiments \cite{Aguilar-Arevalo:2018wea,Aguilar-Arevalo:2017mqx,Dror:2017ehi,NA62:2019meo,Barbieri:1975xy}. The shaded blue region highlights the new parameter space covered by CCM120. Sensitivity projections for the upgraded CCM200 three-year run are shown as colored dotted lines under different assumptions. Blue: current shielding and filtered LAr, orange: improved shielding and filtered LAr, and pink: improved shielding and underground LAr. See Ref.\,\cite{CCM:2021leg} for details.}
    \label{fig:dmCL}
\end{figure}
These are calculated using the same frequentist method used for the MiniBooNE DM and oscillation analyses~\cite{Aguilar-Arevalo:2017mqx,Aguilar-Arevalo:2018wea,Aguilar-Arevalo:2020nvw}.


 
 
 The results of this analysis are displayed in Fig.\,\ref{fig:dmCL}, where CCM120's 90\% C.\,L. median sensitivity and observed limits are shown as dotted black and solid black lines, respectively.
 Previous bounds are shown as shaded gray regions \cite{Aguilar-Arevalo:2018wea,Aguilar-Arevalo:2017mqx,NA62:2019meo,Barbieri:1975xy}, where constraints from UV completions of models with gauged baryon number\,\cite{Dror:2017ehi} are shown only for the most conservative and least model dependent anomalon limit.
 
 Notably, with only 1.5 months of data and contaminated LAr, which reduced the attenuation length of the near UV scintillation light to about 50\,cm, CCM120 is still able to extend the excluded parameter space of this benchmark model beyond the leading constraints previously set by MiniBooNE \cite{Aguilar-Arevalo:2018wea,Aguilar-Arevalo:2017mqx}  and NA62 \cite{NA62:2019meo} within the range $2\;\text{MeV} \lesssim m_\chi \lesssim 29\;\text{MeV}$ (as shown by the shaded blue region).
Furthermore, CCM120's new limit extends the excluded relic density target region from $m_\chi \gtrsim 13$ MeV down to $m_\chi \gtrsim 9$ MeV. CCM120's bounds are even stronger for other slices of parameter space for which $m_{_{V_B}}/m_\chi>2.1$.

\paragraph{\textbf{Conclusions}} CCM120 successfully carried out a search for leptophobic DM using its nuclear scattering data. 
CCM120's greatest success is its ability to use an engineering run with 1.5 months of analyzable data to exclude new parameter space of leptophobic DM models, as illustrated by the shaded blue region in Fig.\,\ref{fig:dmCL} for a benchmark model with scalar DM and a baryonic vector portal. With the improvements already implemented in CCM200 (filtered LAr, more shielding, etc.), a 3-year run collecting $22.5 \times 10^{21}$ POT will significantly extend CCM200's reach to unexplored parameter space of leptophobic DM models with the possibility of completely ruling out the low mass relic density target of the benchmark model considered in this analysis.   


\textbf{\textit{Acknowledgements}} \, We thank Brian Batell for useful discussions on leptophobic DM models. This work was funded by the U.S. Department of Energy, the U.S. National Science Foundation, Los Alamos National Laboratory LDRD program, and PAPIIT-UNAM grant No.\,\,IT100420. We also wish to acknowledge support from the LANSCE Lujan Center and LANL's Accelerator Operations and Technology (AOT) division.  This research used resources provided by the Los Alamos National Laboratory Institutional Computing Program, which is supported by the U.S. Department of Energy National Nuclear Security Administration under Contract No.\,89233218CNA000001.


\bibliography{bibliography.bib}

\providecommand{\noopsort}[1]{}\providecommand{\singleletter}[1]{#1}%
\begin{thebibliography}{18}%
\makeatletter
\providecommand \@ifxundefined [1]{%
 \@ifx{#1\undefined}
}%
\providecommand \@ifnum [1]{%
 \ifnum #1\expandafter \@firstoftwo
 \else \expandafter \@secondoftwo
 \fi
}%
\providecommand \@ifx [1]{%
 \ifx #1\expandafter \@firstoftwo
 \else \expandafter \@secondoftwo
 \fi
}%
\providecommand \natexlab [1]{#1}%
\providecommand \enquote  [1]{``#1''}%
\providecommand \bibnamefont  [1]{#1}%
\providecommand \bibfnamefont [1]{#1}%
\providecommand \citenamefont [1]{#1}%
\providecommand \href@noop [0]{\@secondoftwo}%
\providecommand \href [0]{\begingroup \@sanitize@url \@href}%
\providecommand \@href[1]{\@@startlink{#1}\@@href}%
\providecommand \@@href[1]{\endgroup#1\@@endlink}%
\providecommand \@sanitize@url [0]{\catcode `\\12\catcode `\$12\catcode
  `\&12\catcode `\#12\catcode `\^12\catcode `\_12\catcode `\%12\relax}%
\providecommand \@@startlink[1]{}%
\providecommand \@@endlink[0]{}%
\providecommand \url  [0]{\begingroup\@sanitize@url \@url }%
\providecommand \@url [1]{\endgroup\@href {#1}{\urlprefix }}%
\providecommand \urlprefix  [0]{URL }%
\providecommand \Eprint [0]{\href }%
\providecommand \doibase [0]{https://doi.org/}%
\providecommand \selectlanguage [0]{\@gobble}%
\providecommand \bibinfo  [0]{\@secondoftwo}%
\providecommand \bibfield  [0]{\@secondoftwo}%
\providecommand \translation [1]{[#1]}%
\providecommand \BibitemOpen [0]{}%
\providecommand \bibitemStop [0]{}%
\providecommand \bibitemNoStop [0]{.\EOS\space}%
\providecommand \EOS [0]{\spacefactor3000\relax}%
\providecommand \BibitemShut  [1]{\csname bibitem#1\endcsname}%
\let\auto@bib@innerbib\@empty
\bibitem [{\citenamefont {Batell}\ \emph {et~al.}(2014)\citenamefont {Batell},
  \citenamefont {deNiverville}, \citenamefont {McKeen}, \citenamefont
  {Pospelov},\ and\ \citenamefont {Ritz}}]{Batell:2014yra}%
  \BibitemOpen
  \bibfield  {author} {\bibinfo {author} {\bibfnamefont {B.}~\bibnamefont
  {Batell}}, \bibinfo {author} {\bibfnamefont {P.}~\bibnamefont
  {deNiverville}}, \bibinfo {author} {\bibfnamefont {D.}~\bibnamefont
  {McKeen}}, \bibinfo {author} {\bibfnamefont {M.}~\bibnamefont {Pospelov}},\
  and\ \bibinfo {author} {\bibfnamefont {A.}~\bibnamefont {Ritz}},\ }\bibfield
  {title} {\bibinfo {title} {{Leptophobic Dark Matter at Neutrino Factories}},\
  }\href {https://doi.org/10.1103/PhysRevD.90.115014} {\bibfield  {journal}
  {\bibinfo  {journal} {Phys. Rev.}\ }\textbf {\bibinfo {volume} {D90}},\
  \bibinfo {pages} {115014} (\bibinfo {year} {2014})},\ \Eprint
  {https://arxiv.org/abs/1405.7049} {arXiv:1405.7049 [hep-ph]} \BibitemShut
  {NoStop}%
\bibitem [{\citenamefont {Coloma}\ \emph {et~al.}(2016)\citenamefont {Coloma},
  \citenamefont {Dobrescu}, \citenamefont {Frugiuele},\ and\ \citenamefont
  {Harnik}}]{Coloma:2015pih}%
  \BibitemOpen
  \bibfield  {author} {\bibinfo {author} {\bibfnamefont {P.}~\bibnamefont
  {Coloma}}, \bibinfo {author} {\bibfnamefont {B.~A.}\ \bibnamefont
  {Dobrescu}}, \bibinfo {author} {\bibfnamefont {C.}~\bibnamefont
  {Frugiuele}},\ and\ \bibinfo {author} {\bibfnamefont {R.}~\bibnamefont
  {Harnik}},\ }\bibfield  {title} {\bibinfo {title} {{Dark matter beams at
  LBNF}},\ }\href {https://doi.org/10.1007/JHEP04(2016)047} {\bibfield
  {journal} {\bibinfo  {journal} {JHEP}\ }\textbf {\bibinfo {volume} {04}},\
  \bibinfo {pages} {047}},\ \Eprint {https://arxiv.org/abs/1512.03852}
  {arXiv:1512.03852 [hep-ph]} \BibitemShut {NoStop}%
\bibitem [{\citenamefont {Dror}\ \emph {et~al.}(2017)\citenamefont {Dror},
  \citenamefont {Lasenby},\ and\ \citenamefont {Pospelov}}]{Dror:2017ehi}%
  \BibitemOpen
  \bibfield  {author} {\bibinfo {author} {\bibfnamefont {J.~A.}\ \bibnamefont
  {Dror}}, \bibinfo {author} {\bibfnamefont {R.}~\bibnamefont {Lasenby}},\ and\
  \bibinfo {author} {\bibfnamefont {M.}~\bibnamefont {Pospelov}},\ }\bibfield
  {title} {\bibinfo {title} {{New constraints on light vectors coupled to
  anomalous currents}},\ }\href
  {https://doi.org/10.1103/PhysRevLett.119.141803} {\bibfield  {journal}
  {\bibinfo  {journal} {Phys. Rev. Lett.}\ }\textbf {\bibinfo {volume} {119}},\
  \bibinfo {pages} {141803} (\bibinfo {year} {2017})},\ \Eprint
  {https://arxiv.org/abs/1705.06726} {arXiv:1705.06726 [hep-ph]} \BibitemShut
  {NoStop}%
\bibitem [{\citenamefont {Berlin}\ \emph {et~al.}(2019)\citenamefont {Berlin},
  \citenamefont {Blinov}, \citenamefont {Krnjaic}, \citenamefont {Schuster},\
  and\ \citenamefont {Toro}}]{Berlin:2018bsc}%
  \BibitemOpen
  \bibfield  {author} {\bibinfo {author} {\bibfnamefont {A.}~\bibnamefont
  {Berlin}}, \bibinfo {author} {\bibfnamefont {N.}~\bibnamefont {Blinov}},
  \bibinfo {author} {\bibfnamefont {G.}~\bibnamefont {Krnjaic}}, \bibinfo
  {author} {\bibfnamefont {P.}~\bibnamefont {Schuster}},\ and\ \bibinfo
  {author} {\bibfnamefont {N.}~\bibnamefont {Toro}},\ }\bibfield  {title}
  {\bibinfo {title} {{Dark Matter, Millicharges, Axion and Scalar Particles,
  Gauge Bosons, and Other New Physics with LDMX}},\ }\href
  {https://doi.org/10.1103/PhysRevD.99.075001} {\bibfield  {journal} {\bibinfo
  {journal} {Phys. Rev. D}\ }\textbf {\bibinfo {volume} {99}},\ \bibinfo
  {pages} {075001} (\bibinfo {year} {2019})},\ \Eprint
  {https://arxiv.org/abs/1807.01730} {arXiv:1807.01730 [hep-ph]} \BibitemShut
  {NoStop}%
\bibitem [{\citenamefont {Boyarsky}\ \emph {et~al.}(2021)\citenamefont
  {Boyarsky}, \citenamefont {Mikulenko}, \citenamefont {Ovchynnikov},\ and\
  \citenamefont {Shchutska}}]{Boyarsky:2021moj}%
  \BibitemOpen
  \bibfield  {author} {\bibinfo {author} {\bibfnamefont {A.}~\bibnamefont
  {Boyarsky}}, \bibinfo {author} {\bibfnamefont {O.}~\bibnamefont {Mikulenko}},
  \bibinfo {author} {\bibfnamefont {M.}~\bibnamefont {Ovchynnikov}},\ and\
  \bibinfo {author} {\bibfnamefont {L.}~\bibnamefont {Shchutska}},\ }\bibfield
  {title} {\bibinfo {title} {{Searches for new physics at SND@LHC}},\
  }\href@noop {} {\bibfield  {journal} {\bibinfo  {journal} {{}}\ } (\bibinfo
  {year} {2021})},\ \Eprint {https://arxiv.org/abs/2104.09688}
  {arXiv:2104.09688 [hep-ph]} \BibitemShut {NoStop}%
\bibitem [{\citenamefont {Lisowski}\ and\ \citenamefont
  {Schoenberg}(2006)}]{LISOWSKI2006910}%
  \BibitemOpen
  \bibfield  {author} {\bibinfo {author} {\bibfnamefont {P.~W.}\ \bibnamefont
  {Lisowski}}\ and\ \bibinfo {author} {\bibfnamefont {K.~F.}\ \bibnamefont
  {Schoenberg}},\ }\bibfield  {title} {\bibinfo {title} {{The Los Alamos
  Neutron Science Center}},\ }\href
  {https://doi.org/https://doi.org/10.1016/j.nima.2006.02.178} {\bibfield
  {journal} {\bibinfo  {journal} {Nuclear Instruments and Methods in Physics
  Research Section A: Accelerators, Spectrometers, Detectors and Associated
  Equipment}\ }\textbf {\bibinfo {volume} {562}},\ \bibinfo {pages} {910 }
  (\bibinfo {year} {2006})},\ \bibinfo {note} {proceedings of the 7th
  International Conference on Accelerator Applications}\BibitemShut {NoStop}%
\bibitem [{\citenamefont {Bultman}(1998)}]{Bultman:1998}%
  \BibitemOpen
  \bibfield  {author} {\bibinfo {author} {\bibfnamefont {N.}~\bibnamefont
  {Bultman}},\ }\bibfield  {title} {\bibinfo {title} {{Engineering Design of
  the Lujan Center Target}},\ }in\ \href@noop {} {\emph {\bibinfo {booktitle}
  {ICANS XIV}}}\ (\bibinfo {year} {1998})\ pp.\ \bibinfo {pages}
  {345--360}\BibitemShut {NoStop}%
\bibitem [{\citenamefont {Mocko}\ and\ \citenamefont
  {Muhrer}(2013)}]{MOCKO201327}%
  \BibitemOpen
  \bibfield  {author} {\bibinfo {author} {\bibfnamefont {M.}~\bibnamefont
  {Mocko}}\ and\ \bibinfo {author} {\bibfnamefont {G.}~\bibnamefont {Muhrer}},\
  }\bibfield  {title} {\bibinfo {title} {{Fourth-generation spallation neutron
  target-moderator-reflector-shield assembly at the Manuel Lujan Jr. neutron
  scattering center}},\ }\href
  {https://doi.org/https://doi.org/10.1016/j.nima.2012.11.103} {\bibfield
  {journal} {\bibinfo  {journal} {Nuclear Instruments and Methods in Physics
  Research Section A: Accelerators, Spectrometers, Detectors and Associated
  Equipment}\ }\textbf {\bibinfo {volume} {704}},\ \bibinfo {pages} {27 }
  (\bibinfo {year} {2013})}\BibitemShut {NoStop}%
\bibitem [{\citenamefont {Dobrescu}\ and\ \citenamefont
  {Frugiuele}(2015)}]{Dobrescu:2014ita}%
  \BibitemOpen
  \bibfield  {author} {\bibinfo {author} {\bibfnamefont {B.~A.}\ \bibnamefont
  {Dobrescu}}\ and\ \bibinfo {author} {\bibfnamefont {C.}~\bibnamefont
  {Frugiuele}},\ }\bibfield  {title} {\bibinfo {title} {{GeV-Scale Dark Matter:
  Production at the Main Injector}},\ }\href
  {https://doi.org/10.1007/JHEP02(2015)019} {\bibfield  {journal} {\bibinfo
  {journal} {JHEP}\ }\textbf {\bibinfo {volume} {02}},\ \bibinfo {pages}
  {019}},\ \Eprint {https://arxiv.org/abs/1410.1566} {arXiv:1410.1566 [hep-ph]}
  \BibitemShut {NoStop}%
\bibitem [{\citenamefont {deNiverville}\ \emph {et~al.}(2015)\citenamefont
  {deNiverville}, \citenamefont {Pospelov},\ and\ \citenamefont
  {Ritz}}]{deNiverville:2015mwa}%
  \BibitemOpen
  \bibfield  {author} {\bibinfo {author} {\bibfnamefont {P.}~\bibnamefont
  {deNiverville}}, \bibinfo {author} {\bibfnamefont {M.}~\bibnamefont
  {Pospelov}},\ and\ \bibinfo {author} {\bibfnamefont {A.}~\bibnamefont
  {Ritz}},\ }\bibfield  {title} {\bibinfo {title} {{Light new physics in
  coherent neutrino-nucleus scattering experiments}},\ }\href
  {https://doi.org/10.1103/PhysRevD.92.095005} {\bibfield  {journal} {\bibinfo
  {journal} {Phys. Rev.}\ }\textbf {\bibinfo {volume} {D92}},\ \bibinfo {pages}
  {095005} (\bibinfo {year} {2015})},\ \Eprint
  {https://arxiv.org/abs/1505.07805} {arXiv:1505.07805 [hep-ph]} \BibitemShut
  {NoStop}%
\bibitem [{\citenamefont {Aguilar-Arevalo}\ \emph
  {et~al.}(2021{\natexlab{a}})\citenamefont {Aguilar-Arevalo} \emph
  {et~al.}}]{CCM:2021leg}%
  \BibitemOpen
  \bibfield  {author} {\bibinfo {author} {\bibfnamefont {A.~A.}\ \bibnamefont
  {Aguilar-Arevalo}} \emph {et~al.} (\bibinfo {collaboration} {CCM}),\
  }\href@noop {} {\bibinfo {title} {{First Dark Matter Search Results From
  Coherent CAPTAIN-Mills}}} (\bibinfo {year} {2021}{\natexlab{a}}),\ \Eprint
  {https://arxiv.org/abs/2105.14020v2} {arXiv:2105.14020v2 [hep-ex]}
  \BibitemShut {NoStop}%
\bibitem [{\citenamefont {Aguilar-Arevalo}\ \emph {et~al.}(2018)\citenamefont
  {Aguilar-Arevalo} \emph {et~al.}}]{Aguilar-Arevalo:2018wea}%
  \BibitemOpen
  \bibfield  {author} {\bibinfo {author} {\bibfnamefont {A.~A.}\ \bibnamefont
  {Aguilar-Arevalo}} \emph {et~al.} (\bibinfo {collaboration} {MiniBooNE DM}),\
  }\bibfield  {title} {\bibinfo {title} {{Dark Matter Search in Nucleon, Pion,
  and Electron Channels from a Proton Beam Dump with MiniBooNE}},\ }\href
  {https://doi.org/10.1103/PhysRevD.98.112004} {\bibfield  {journal} {\bibinfo
  {journal} {Phys. Rev. D}\ }\textbf {\bibinfo {volume} {98}},\ \bibinfo
  {pages} {112004} (\bibinfo {year} {2018})},\ \Eprint
  {https://arxiv.org/abs/1807.06137} {arXiv:1807.06137 [hep-ex]} \BibitemShut
  {NoStop}%
\bibitem [{\citenamefont {Aguilar-Arevalo}\ \emph {et~al.}(2017)\citenamefont
  {Aguilar-Arevalo} \emph {et~al.}}]{Aguilar-Arevalo:2017mqx}%
  \BibitemOpen
  \bibfield  {author} {\bibinfo {author} {\bibfnamefont {A.~A.}\ \bibnamefont
  {Aguilar-Arevalo}} \emph {et~al.} (\bibinfo {collaboration} {MiniBooNE}),\
  }\bibfield  {title} {\bibinfo {title} {{Dark Matter Search in a Proton Beam
  Dump with MiniBooNE}},\ }\href
  {https://doi.org/10.1103/PhysRevLett.118.221803} {\bibfield  {journal}
  {\bibinfo  {journal} {Phys. Rev. Lett.}\ }\textbf {\bibinfo {volume} {118}},\
  \bibinfo {pages} {221803} (\bibinfo {year} {2017})},\ \Eprint
  {https://arxiv.org/abs/1702.02688} {arXiv:1702.02688 [hep-ex]} \BibitemShut
  {NoStop}%
\bibitem [{\citenamefont {deNiverville}\ \emph {et~al.}(2017)\citenamefont
  {deNiverville}, \citenamefont {Chen}, \citenamefont {Pospelov},\ and\
  \citenamefont {Ritz}}]{deNiverville:2016rqh}%
  \BibitemOpen
  \bibfield  {author} {\bibinfo {author} {\bibfnamefont {P.}~\bibnamefont
  {deNiverville}}, \bibinfo {author} {\bibfnamefont {C.-Y.}\ \bibnamefont
  {Chen}}, \bibinfo {author} {\bibfnamefont {M.}~\bibnamefont {Pospelov}},\
  and\ \bibinfo {author} {\bibfnamefont {A.}~\bibnamefont {Ritz}},\ }\bibfield
  {title} {\bibinfo {title} {{Light dark matter in neutrino beams: production
  modelling and scattering signatures at MiniBooNE, T2K and SHiP}},\ }\href
  {https://doi.org/10.1103/PhysRevD.95.035006} {\bibfield  {journal} {\bibinfo
  {journal} {Phys. Rev.}\ }\textbf {\bibinfo {volume} {D95}},\ \bibinfo {pages}
  {035006} (\bibinfo {year} {2017})},\ \Eprint
  {https://arxiv.org/abs/1609.01770} {arXiv:1609.01770 [hep-ph]} \BibitemShut
  {NoStop}%
\bibitem [{\citenamefont {Cuesta}(2021)}]{ProtoDUNE:2021cuesta}%
  \BibitemOpen
  \bibfield  {author} {\bibinfo {author} {\bibfnamefont {C.}~\bibnamefont
  {Cuesta}} (\bibinfo {collaboration} {ProtoDUNE}),\ }\href@noop {} {\bibinfo
  {title} {{Scintillation light detection in the long-drift ProtoDUNE-DP liquid
  argon TPC}}} (\bibinfo {year} {2021}),\ \Eprint
  {https://arxiv.org/abs/2106.15334v1} {arXiv:2106.15334v1 [physics.ins-det]}
  \BibitemShut {NoStop}%
\bibitem [{\citenamefont {Cortina~Gil}\ \emph {et~al.}(2019)\citenamefont
  {Cortina~Gil} \emph {et~al.}}]{NA62:2019meo}%
  \BibitemOpen
  \bibfield  {author} {\bibinfo {author} {\bibfnamefont {E.}~\bibnamefont
  {Cortina~Gil}} \emph {et~al.} (\bibinfo {collaboration} {NA62}),\ }\bibfield
  {title} {\bibinfo {title} {{Search for production of an invisible dark photon
  in $\pi^0$ decays}},\ }\href {https://doi.org/10.1007/JHEP05(2019)182}
  {\bibfield  {journal} {\bibinfo  {journal} {JHEP}\ }\textbf {\bibinfo
  {volume} {05}},\ \bibinfo {pages} {182}},\ \Eprint
  {https://arxiv.org/abs/1903.08767} {arXiv:1903.08767 [hep-ex]} \BibitemShut
  {NoStop}%
\bibitem [{\citenamefont {Barbieri}\ and\ \citenamefont
  {Ericson}(1975)}]{Barbieri:1975xy}%
  \BibitemOpen
  \bibfield  {author} {\bibinfo {author} {\bibfnamefont {R.}~\bibnamefont
  {Barbieri}}\ and\ \bibinfo {author} {\bibfnamefont {T.~E.~O.}\ \bibnamefont
  {Ericson}},\ }\bibfield  {title} {\bibinfo {title} {{Evidence Against the
  Existence of a Low Mass Scalar Boson from Neutron-Nucleus Scattering}},\
  }\href {https://doi.org/10.1016/0370-2693(75)90073-8} {\bibfield  {journal}
  {\bibinfo  {journal} {Phys. Lett. B}\ }\textbf {\bibinfo {volume} {57}},\
  \bibinfo {pages} {270} (\bibinfo {year} {1975})}\BibitemShut {NoStop}%
\bibitem [{\citenamefont {Aguilar-Arevalo}\ \emph
  {et~al.}(2021{\natexlab{b}})\citenamefont {Aguilar-Arevalo} \emph
  {et~al.}}]{Aguilar-Arevalo:2020nvw}%
  \BibitemOpen
  \bibfield  {author} {\bibinfo {author} {\bibfnamefont {A.~A.}\ \bibnamefont
  {Aguilar-Arevalo}} \emph {et~al.} (\bibinfo {collaboration} {MiniBooNE}),\
  }\bibfield  {title} {\bibinfo {title} {{Updated MiniBooNE neutrino
  oscillation results with increased data and new background studies}},\ }\href
  {https://doi.org/10.1103/PhysRevD.103.052002} {\bibfield  {journal} {\bibinfo
   {journal} {Phys. Rev. D}\ }\textbf {\bibinfo {volume} {103}},\ \bibinfo
  {pages} {052002} (\bibinfo {year} {2021}{\natexlab{b}})},\ \Eprint
  {https://arxiv.org/abs/2006.16883} {arXiv:2006.16883 [hep-ex]} \BibitemShut
  {NoStop}%
\end{thebibliography}%


\providecommand{\noopsort}[1]{}\providecommand{\singleletter}[1]{#1}%
%

\end{document}